%% file: paper.tex
\documentclass[preprint]{aastex502perso}
\usepackage[T1]{fontenc}
\usepackage{amsmath}
\usepackage[english]{babel}
\usepackage{graphics}

\usepackage{amssymb}

\begin{document}

\renewcommand{\baselinestretch}{1.12}

\input com.tex
\def\ga{\mathrel{\mathchoice {\vcenter{\offinterlineskip\halign{\hfil
$\displaystyle##$\hfil\cr>\cr\sim\cr}}}
{\vcenter{\offinterlineskip\halign{\hfil$\textstyle##$\hfil\cr
>\cr\sim\cr}}}
{\vcenter{\offinterlineskip\halign{\hfil$\scriptstyle##$\hfil\cr
>\cr\sim\cr}}}
{\vcenter{\offinterlineskip\halign{\hfil$\scriptscriptstyle##$\hfil\cr
>\cr\sim\cr}}}}}

\bibliographystyle{plain}
\slugcomment{Accepted by Astrophysical Journal Letters, 
26 June 2002}

\title{Angular momentum extraction by gravity waves in the Sun}
\author{Suzanne Talon}
\affil{D\'epartement de Physique, Universit\'e de Montr\'eal, Montr\'eal PQ H3C 3J7}
\affil{CERCA, 5160, boul. D\'ecarie, suite 400, Montr\'eal PQ H3X 2H9}
\author{Pawan Kumar}
\affil{Dept. of Astronomy, University of Texas, Austin, TX 78713}
\author{Jean-Paul Zahn}
\affil{LUTH, Observatoire de Paris, 92195 Meudon, France}

\begin{abstract}
We review the behavior of the oscillating shear layer produced
by gravity waves below the surface convection zone of the Sun.
We show that, under asymmetric filtering produced by
this layer, gravity waves of low spherical order,
which are stochastically excited at the base of the convection
zone of late type stars, can extract angular momentum from their 
radiative interior. The time-scale for this momentum
extraction in a Sun-like star is of the order of $10^7$ years.
The process is particularly efficient in the central region,
and it could produce there a slowly rotating core.

\end{abstract}


\newpage\section{Introduction} 

Angular momentum transport by gravity waves has received much attention
recently. While it has first been considered as a key mechanism
in the tidal interaction of binary systems (Zahn 1975; Goldreich
\& Nicholson 1989), it has since been invoked also as an efficient process
of momentum redistribution in single stars (Schatzman 1993;
Kumar \& Quataert 1997; Zahn, Talon \& Matias 1997). 
The mechanism proposed was similar to that acting
in binary stars, with synchronization  proceeding gradually inwards.
However, as pointed out by Gough \& McIntyre (1998) and Ringot (1998), the
treatment of angular momentum extraction as presented
by these first studies was
incorrect, and it became clear that the actual properties
of wave transport were
far more complex than originally anticipated.

Wave properties have been examined further by Barnes, McGregor \& 
Charbonneau (1998) who showed that magnetic fields transform pure
gravity waves into gravito-Alfv\'en waves, thus modifying
their damping behavior. They showed that strong magnetic field may
prevent certain waves from propagating.

Kumar, Talon \& Zahn (1999, hereafter KTZ) then demonstrated how gravity waves in 
interaction
with shear turbulence may lead to the formation of an oscillating shear
layer just below the surface convection zone, 
analogous to the well studied quasi-biennial oscillation
of atmospheric sciences (see \eg Shepherd 2000 for details on the QBO
and other wave-driven oscillations).

Kim \& McGregor (2001) further studied that oscillation in a simplified
two waves model. They showed that, with only a prograde and
a retrograde wave, and with the velocity fixed both at the top and
at the bottom of the solar tachocline, the time-dependent behavior
of the rotation profile goes from being periodic, to quasi-periodic and to
chaotic as the viscosity in the shear region is slowly decreased.
Their goal was to seek an explanation of a 1.3 yr variation,
which Howe et al. (2000) claim to have detected in the tachocline.

Here we want to reexamine the effect of waves in the deep solar
interior. Indeed, while the high degree waves are damped very close
to the base of the convection zone, 
thus leading to rapid oscillations that could
explain rotational velocity variations in the tachocline 
(Kim \& McGregor 2001) or be related to the solar cycle (KTZ), the low 
degree waves that are also excited stochastically by the
convective motions, although with a somewhat lower efficiency 
(see Kumar \& Quataert 1997 and 
Zahn, Talon \& Matias 1997) will have, over a much longer period,
an effect on the deep interior.

In this letter, we wish to demonstrate how these waves, in conjunction with
shear turbulence, indeed achieve
momentum extraction in the deep interior.
In \S \ref{shme}, we explain heuristically how that process
takes places and in \S \ref{nums}, we show the results
of a long numerical simulation applied to the solar case.

\begin{figure}[t]
{\centering \resizebox*{0.46 \textwidth}{!}
{\rotatebox{0}{\includegraphics{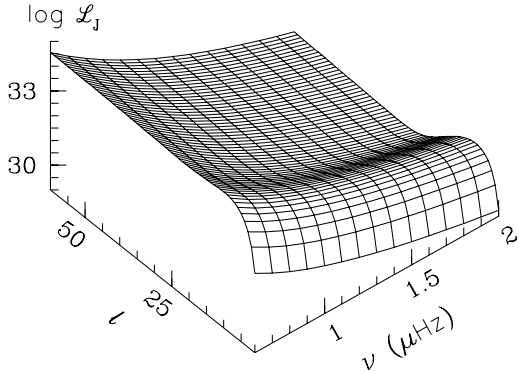}}} \par}
\caption{Angular momentum luminosity integrated over $0.1 \mu{\rm Hz}$
as a function of order $\ell$ and frequency.
\label{fig:mom_lum}}
\end{figure}


\begin{figure}[t]
{\centering \resizebox*{0.46 \textwidth}{!}
{\rotatebox{-90}{\includegraphics{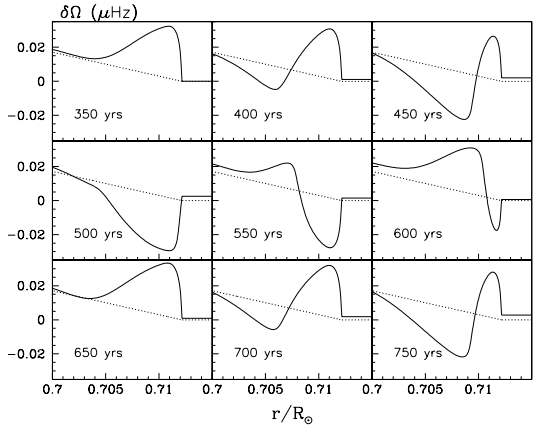}}} \par}
\caption{Oscillating shear layer below the surface convection zone.
The dotted line shows the initial rotation profile.
With the surface rotating slower than the core, a prograde shear
layer is initially formed, followed by a retrograde one. When the
shear becomes too intense, turbulent viscosity acts to merge the
prograde layer with the convection zone, leaving behind the retrograde layer. 
A new prograde layer forms behind, and migrates towards the convection zone when the
retrograde layer is  absorbed. The cycle then resumes.
\label{fig:local}}
\end{figure}


\begin{figure}[t]
{\centering \resizebox*{0.46 \textwidth}{!}
{\rotatebox{0}
{\includegraphics{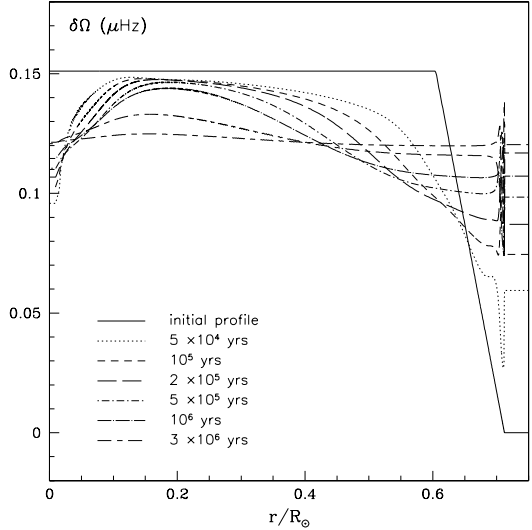}}} \par}
\caption{Global evolution of the rotation profile over large time
periods. Efficient momentum extraction is visible in a characteristic
time-scale of $10^5$ years. \label{fig:global}}
\end{figure}


\begin{table}[t]
\caption{Damping length $d_\omega$  (scaled by the solar radius)
at $0.6 ~ \mu {\rm Hz}$}
\begin{minipage}{\linewidth}
\begin{tabular}{rrrr}\hline
\multicolumn{1}{c}{$\ell$} &
\multicolumn{1}{c}{$m$} &
\multicolumn{1}{c}{$d_\omega$\footnote{In solid body rotation}} &
\multicolumn{1}{c}{$d_\omega$\footnote{With the initial rotation profile 
(\cf Figures~\ref{fig:local} \&~\ref{fig:global})}} \\
\hline
\hline
 1  &  1  & 0.616 & 0.461 \\
    & -1  &       & 0.661\vspace{0.1cm}\\
 5  &  5  & 0.084 & 0.036 \\
    & -5  &       & 0.499\vspace{0.1cm}\\
 9  &  9  & 0.033 & 0.017 \\
    & -9  &       & 0.378\vspace{0.1cm}\\
13  &  13 & 0.018 & 0.010 \\
    & -13 &       & 0.256\vspace{0.1cm}\\
17  &  17 & 0.012 & 0.007 \\
    & -17 &       & 0.061\vspace{0.1cm}\\
21  &  21 & 0.008 & 0.005 \\
    & -21 &       & 0.024\vspace{0.1cm}\\
\end{tabular}
\end{minipage}\label{tab:damp}
\end{table}

\section{Shear layer and momentum extraction \label{shme}}

The formation of a shear layer is easy to describe, using a simple
two waves model. When both a prograde and a retrograde wave travel
in an accelerating region (\ie where $\Omega$ increases inwards with depth), 
the prograde wave frequency diminishes,
enhancing its damping; the opposite occurs for the retrograde wave (see
KTZ for details). This creates a double peaked shear layer.
The gradient in that layer rises until it becomes steep enough
so that turbulence will start playing a role and will effectively
merge the shear layer with the adjacent convection zone.
The retrograde layer\footnote{The retrograde layer is the one that rotates
more slowly than the convection zone, while the prograde layer rotates more
rapidly.} will then slowly rise toward the convection zone,
and a prograde layer will form further inside the star. The retrograde
layer is eventually also absorbed by the convection zone, and the cycle
goes on.

The oscillating shear layer will play an important role in filtering the 
(low degree) waves
that will penetrate, and thus deposit their momentum in the
deep interior. Indeed, when the prograde peek is larger, it is the
retrograde waves that will penetrate and deposit their negative
momentum in the interior,
and vice versa if the retrograde peek dominates. If there is
no initial differential rotation in the interior, the shear layer
will oscillate symmetrically, and on average no net 
momentum flux will occur. However, if
the surface convection zone is slowed down, as in the case of the Sun,
there exists a bias in the formation of the shear layer, which
will favor the prograde peek (the interior then rotating faster
than the surface), thus letting a larger amount of negative momentum
being transported to the deep interior. The positive momentum
deposited in excess in the shear layer will then be transfered back in
the surface convection zone by turbulence. The net effect will thus
be to extract momentum from the radiation zone.

\section{A numerical simulation \label{nums}}

While it is clear that the filtering bias introduced by the slope
of the rotation profile below the convection zone
leads to momentum extraction, we further need to know where exactly
this extraction takes place, and how efficient it is.  
To answer this question, we performed a numerical
simulation covering a very long time period. 
There are two key ingredients in such a simulation.
Firstly, one needs to include a whole variety of radial orders
to describe both the waves responsible for the formation and
oscillation of the shear layer and the deeply penetrating waves.
Furthermore, retaining only two wave orders has the effect of depositing
momentum too locally, compared to having a more complete spectrum.
We thus included spherical orders $\ell=1,5,9,...,57$, to capture
a wide variety of damping behaviors. We also considered a small frequency range
$\omega = 0.6$ to $1.0 ~\mu{\rm Hz}$. 
Higher frequency waves will have little impact since the
power spectrum we are using in these simulations falls off 
as $\omega^{-4.5}$ (\cf Eq. \ref{gold} and KTZ), while lower
frequency waves will be damped very close to the convection zone 
(\cf Eq. \ref{optdepth}), not undergoing differential damping.
The second key ingredient is to follow both 
short time-scales ($\sim  10$ years), characteristic of the shear layer
oscillation, and long time-scales ($\sim 10^7$ years),
characteristic of the spin-down rate. This combinaison requires the use
of a conservative numerical scheme.

We thus calculated the evolution of angular momentum in the radiative interior
by the combined effect of turbulence and gravity waves:
\begin{equation}
\rho \dtt \lc r^2 {\Omega}\rc = \frac{1}{ r^2} \drr \lc \rho \nu_t r^4 \dr{\Omega} \rc 
- \frac{3}{8\pi} \frac{1}{r^2} \drr{{\cal L}_J(r)} 
\label{ev_omega}
\end{equation}
where $\rho$ is the density, $\Omega$ the angular velocity and
$\nu_t$ the turbulent viscosity (\cf Talon 1997).
The local momentum luminosity ${\cal L}_J(r)$ is integrated over
the whole wave spectrum, each wave being damped by radiative processes
and turbulent viscosity (see Eq. \ref{turbvisc})
\begin{equation}
{\cal L}_J(r) = \sum_{\sigma, \ell, m} {{\cal L}_J}_{\ell, m} \lp r_{\rm zc}\rp
\exp \lc -\tau(r, \sigma, \ell) \rc
\end{equation}
and where the local damping rate takes into account the mean molecular weight
stratification
\begin{equation}
\tau(r, \sigma, \ell) = [\ell(\ell+1)]^{3\over2} \int_r^{r_c} 
\lp K + \nu_t \rp \; {N N_T^2 \over
\sigma^4}  \left({N^2 \over N^2 - \sigma^2}\right)^{1 \over 2} {\diff r
\over r^3} \label{optdepth}
\end{equation}
where $N^2 = N_T^2 + N_{\mu}^2$ is the total Brunt-V\"ais\"al\"a frequency,
$N_T^2$ is its thermal part and $N_{\mu}^2$ is due to the
mean molecular weight stratification (\cf Zahn et al. 1997).
$\sigma(r,m) = \omega - m \lc\Omega(r)-\Omega_{\rm zc}\rc $ is the
local (Doppler shifted) frequency, and $\omega$ is the frequency in the reference
frame of the convection zone.
The damping length $d_\omega$ defined by $\tau(d_\omega) = 1$ depends strongly
on frequency $\omega$ and, in presence of differential rotation, 
on the signed value of the azimuthal wavenumber $m$, 
as shown in Table \ref{tab:damp}.
The evolution equation (\ref{ev_omega}) 
is solved using a decoupled scheme, in which momentum is first deposited
by waves and 
diffusion is then treated using a perfectly conservative finite element method.
After $10^7$ time steps of 1 year, the error on 
global momentum conservation is $\sim 10^{-5}$.
This equation is solved throughout the radiation zone.
The upper boundary condition expresses the conservation of
momentum of the star as a whole, with the convection zone rotating 
as a solid body. 
On their way to the core, the waves are reflected when their
local frequency $\sigma$ equals the Brunt-V\"ais\"al\"a frequency 
\begin{equation}
\sigma(r,m) = N.
\end{equation}
They then deposit more momentum as they travel back to the convection zone.
Some waves do travel all the way to the core. Since that
region contains very little momentum, it is easily spun down
(see Figure \ref{fig:global}).
Waves reaching the center most region (below $r=0.01~R_\odot$)
are reflected back; 
this innermost core has been taken to rotate as a solid body.

Wave generation is due to Reynolds stresses at the base of the convection zone,
and to model its amplitude we followed here the description
of Goldreich et al. (1994) (see also KTZ). The energy flux per 
unit frequency ${\cal F}_E$ is then
\begin{eqnarray}
{\cal F}_E \lp \ell, \omega \rp &=& \frac{\omega^2}{4\pi} \int dr\; \frac{\rho^2}{r^2} 
   \left[\left(\frac{\partial \xi_r}{\partial r}\right)^2 + 
   \ell(\ell+1)\left(\frac{\partial \xi_h}{\partial r}\right)^2 \right]  \nonumber \\
 && \times  \exp\left[ -h_\omega^2 \ell(\ell+1)/2r^2\right] \frac{v^3 L^4 }{1 
  + (\omega \tau_L)^{15/2}},
\label{gold}
\end{eqnarray}
where 
$\xi_r$ and $[\ell(\ell+1)]^{1/2}\xi_h$ are the radial and horizontal
displacement wavefunctions which are normalized to unit energy flux just 
below the convection zone, $v$ is the convective velocity, $L$ is the radial
size of an energy bearing turbulent eddy, $\tau_L \approx L/v$ is the
characteristic convective time, and $h_\omega$ is the
radial size of the largest eddy at $r$ with characteristic frequency of
$\omega$ or greater ($h_\omega = L \min\{1, (2\omega\tau_L)^{-3/2}\}$).
The gravity waves are evanescent in the convection zone, the region
where they are excited.
The above equation was derived under the assumption that the
turbulence spectrum is Kolmogorov and
it ignores wave excitation resulting from convective overshooting.
The momentum flux per unit frequency ${\cal F}_J$ is then
\begin{equation}
{{\cal F}_J}\lp m, \ell, \omega \rp = \frac{m}{\omega} {\cal F}_E\lp \ell, \omega \rp.
\end{equation}

In the case of the Sun, there is a whole spectrum of waves,
in frequency $\omega$ and spherical order $\ell$.
Their range is given by $\omega \leq N_c$, 
where $N_c$ is the Brunt-V\"ais\"al\"a
frequency at the base of the convection zone\footnote{This
frequency depends on the amount of overshooting that is considered.
However, since the power spectrum scales as $\omega^{-4.5}$ (\cf KTZ),
its exact value is not crucial.}, and by
$1 \leq \ell \leq \ell_c$ where $\ell_c \sim 60$ is the spherical 
order characterizing convection (it corresponds to 
one pressure scale-height). Assuming that all azimuthal orders $m$ are equally excited,
we find that most of the momentum is carried by the low frequency
waves. There is significant momentum luminosity in
the low order waves that do penetrate deep in the interior (\cf Figure ~\ref{fig:mom_lum}
and Table \ref{tab:damp}).

For this first exploration, turbulent viscosity has been
taken proportional to the radiative
viscosity ($\nu_{\rm rad}$), admittedly a very crude prescription:
\begin{equation}
\nu_t = 10^5 \nu_{\rm rad} = 10^5 \frac{\chi T}{5c^2 \rho} \label{turbvisc}
\end{equation}
where $\chi$ is the thermal conductivity and $c$ the speed of light.
The total energy flux in the considered waves is $6.3 \times 10^{29} {~\rm
erg/s}$, and the total integrated momentum luminosity is
$1.7 \times 10^{36} {~\rm g \,cm^2 /s^2}$.
This turbulent viscosity operates both on the mean
rotation profile and on the waves (\cf Eq. \ref{optdepth} and \ref{ev_omega}).

This simulation is performed in one dimension and we only consider 
the effect of spherically integrated waves. We thus neglect the
latitudinal differential rotation just below the convection zone, as
well as the influence of the Coriolis force on wave propagation
and wave excitation.

The results are displayed in Figures \ref{fig:local} and \ref{fig:global}.
The first shows the shear layer oscillation cycle, which occurs here
with a period of $\sim 300 {\rm ~years}$. The second shows
the evolution of the rotation profile
in the deep interior over a much longer period.
Momentum extraction is related to the asymmetric filtering by the double shear
layer, which is determined by the
amount of differential rotation initially present below the convection zone.
In our simulation we let $\Omega$ increase by $0.15 ~{\rm \mu Hz}$
down to $r=0.60 R_\odot$.
This profile, applied to the wave spectrum we consider here,
yields a differential momentum flux below the shear layer of
$ 2 \times 10^{33} {~\rm g\, cm^2 /s^2}$. In $5 \times 10^4$ years,
this leads to an extraction of $2.5 \times 10^{45} {~\rm g\, cm^2/s}$, in good
agreement with the numerical simulation which shows an extraction
of $4 \times 10^{45} {~\rm g\, cm^2/s}$.

As the simulation progresses,
the radiation zone decelerates and the
convection zone accelerates\footnote{In an evolving solar model, 
the surface convection zone would constantly
be spinned down by a magnetic torque and would thus probably never
accelerate.}; this leads to less differential 
filtering and therefore momentum extraction gradually slows down.
The simulation shows that waves are very efficient in extracting momentum
from the deep core, in a time short compared to the main-sequence life time.
At later times, the net wave flux decreases, allowing turbulence to
smooth the rotation profile.  
But the turbulent transport will also taper off, 
due to the build-up of a stabilizing helium gradient.
Therefore it is plausible that in the present Sun
the central region is rotating
slower than the rest of the radiation zone,
as suggested by certain helioseismic inversions 
(Elsworth et al. 1995; Corbard et al. 1997). 
If this trend is confirmed, it would imply
that wave transport is more efficient than magnetic torquing, 
which has also been invoked for the extraction of angular momentum
(\cf Mestel \& Weiss 1987, Charbonneau \& MacGregor 1993).

\section{Discussion}

In this letter, we have shown how gravity waves can extract
angular momentum from the radiative core of a solar-like star through
differential wave filtering, sufficiently to establish an 
almost uniform rotation profile in about $10^7$ years. 
In our model, the presence of turbulent transport
is essential to yield the
proper result; indeed, it is required both to produce the QBO like
oscillation, and to smooth the excessive differential rotation
produced by waves in the central region. 

The evolution of latitudinally averaged angular momentum 
in the Sun would then proceed as follows. The magnetic torque applied 
to the convection zone tends to slow it down,
establishing a negative gradient of angular velocity, which induces 
the bias between prograde and retrograde waves. Due to the combined 
action of wave transport and of turbulent viscosity, the velocity gradient 
tends to diminish. In our simulation it disappears altogether, but
in the real Sun, it would be maintained at some level due to the
competition between the extraction of angular momentum from the
radiative interior, and its loss by the solar wind: the faster
the spin-down, the steeper the gradient, the stronger the extraction. 
If the extraction of angular momentum by gravity waves is as 
efficient as we estimate, it will adjust itself such as to provide
just the flux of angular momentum which is lost by the wind,
with the interior profile of angular velocity being rather flat,
except near the center and at the top of the radiation zone.

The presence of a large toroidal magnetic at the base of the convection
zone could somewhat lengthen the time-scales mentioned here.
Indeed, the magnetic field required to
prevent wave propagation at $0.6 ~{\rm \mu Hz}$ is about
$(3 \times 10^5/\ell)~{\rm G}$ (\cf KTZ). If fields of a strength as
high as $10^5 ~{\rm G}$ are present in that region (\cf
Fan et al. 1993, Caligari et al. 1995), high $\ell$ modes could
be prevented from propagating, modifying the evolution of the
double shear layer. However, since such large fields
occur in localized areas, the global effect would rather be one
of decreasing the wave flux and thus, simply increase the time
scale estimated here.

This initial work must be pursued in many ways. Firstly, one
would like to assess the efficiency of this mechanism
in an evolving solar model, with magnetic spin-down, 
in which meridional circulation would also be taken into account
in order to verify the scenario we proposed earlier.

Other improvements should be made on the physical description.
The effect of the Coriolis force must be included, which 
could require to step up to two-dimensional simulations.
Finally, a more realistic prescription for the turbulent transport must be
implemented, before we can conclude that the solar core is
rotating significantly slower than the rest of the radiation zone.

\acknowledgments
{\noindent} {\bf Acknowledgments}  ~S. Talon 
was supported by NSERC of Canada and by the Canada Research
Chair in Stellar Astrophysics awarded to Prof. G.~Fontaine.


\end{document}

%% file: com.tex
\def\etal{et al.\ }
\def\chaphead{}
\def\absmath{\textfont0=\eightrm \scriptfont0=\sixrm
	      \textfont1=\eightmit \scriptfont1=\sixmit}
\def\absfont{\let\rm=\eightrm \let\it=\eightit \rm\absmath}
\def\regmath{\textfont0=\twelverm \scriptfont0=\tenrm
	      \textfont1=\twelvemit \scriptfont1=\tenmit}
\def\peterfont{\let\rm=\twelverm \let\it=\twelveit \rm\regmath}
\def\b#1{\skew{-2}\vec#1}  

\def\emp{\sl}
\def\deffn{\bf}
\def\deg{^\circ}
\def\Vlasov{collisionless Boltzmann\ }
\def\lsls{\ll}
\def\grgr{\gg}
\def\erf{\mathop{\rm erf}\nolimits} 
\def\eqv{\equiv}
\def\real{\Re e}
\def\imag{\Im m}
\def\ctrline#1{\centerline{#1}}
\def\spose#1{\hbox to 0pt{#1\hss}}
\def\s{\ifmmode \widetilde \else \~\fi} 
     
\newcount\notenumber
\notenumber=1
\newcount\eqnumber
\eqnumber=1
\newcount\fignumber
\fignumber=1
     
\def\yyskip{\penalty-100\vskip6pt plus6pt minus4pt}
     
\def\numberpara{\yyskip\noindent}
     
\def\km{{\rm\,km}}
\def\kms{{\rm\,km\,s^{-1}}}
\def\kpc{{\rm\,kpc}}
\def\mpc{{\rm\,Mpc}}
\def\msun{{\rm\,M_\odot}}
\def\lsun{{\rm\,L_\odot}}
\def\rsun{{\rm\,R_\odot}}
\def\pc{{\rm\,pc}}
\def\cm{{\rm\,cm}}
\def\yr{{\rm\,yr}}
\def\au{{\rm\,AU}}
\def\AU{{\rm\,AU}}
\def\gm{{\rm\,g}}
\def\s{{\rm\,s}}
\def\dyne{{\rm\,dyne}}
\def\G{{\rm\,G}}
\def\erg{{\rm\,erg}}
\def\K{{\rm\, K}}
\def\rearth{{\rm\,R_\oplus}}
\def\mearth{{\rm\,M_\oplus}}

\def\note#1{\footnote{$^{\the\notenumber}$}{#1}\global\advance\notenumber by 1}
\def\foot#1{\raise3pt\hbox{\eightrm \the\notenumber}
     \hfil\par\vskip3pt\hrule\vskip6pt
     \noindent\raise3pt\hbox{\eightrm \the\notenumber}
     #1\par\vskip6pt\hrule\vskip3pt\noindent\global\advance\notenumber by 1}
\def\propo{\propto}
\def\larrow{\leftarrow}
\def\rarrow{\rightarrow}
          
\def\Dt{\spose{\raise 1.5ex\hbox{\hskip3pt$\mathchar"201$}}}    
\def\dt{\spose{\raise 1.0ex\hbox{\hskip2pt$\mathchar"201$}}}    
\def\llangle{\langle\langle}
\def\rrangle{\rangle\rangle}
\def\ldotss{\ldots}
\def\del{\b\nabla}
     
\def\refindent{\par\noindent\parskip=4pt\hangindent=3pc\hangafter=1 }
\def\apj#1#2#3{\refindent#1.  {\sl Astrophys.\  J. }{\bf#2}, #3.}
\def\apjlett#1#2#3{\refindent#1.  {\sl Astrophys.\  J. Lett. }{\bf#2}, #3.}
\def\mn#1#2#3{\refindent#1.  {\sl Mon. Not. Roy. Astron. Soc. }{\bf#2}, #3.}
\def\mnras#1#2#3{\refindent#1.  {\sl Mon. Not. Roy. Astron. Soc. }{\bf#2}, #3.}
\def\aj#1#2#3{\refindent#1.  {\sl Astron. J. }{\bf#2}, #3.}
\def\aa#1#2#3{\refindent#1.  {\sl Astron. Astrophys. }{\bf#2}, #3.}
\def\Nature#1#2#3{\refindent#1.  {\sl Nature }{\bf#2}, #3.}
\def\Icarus#1#2#3{\refindent#1.  {\sl Icarus }{\bf#2}, #3.}
\def\refpaper#1#2#3#4{\refindent#1.  {\sl #2 }{\bf#3}, #4.}
\def\refbook#1{\refindent#1}
\def\science#1#2#3{\refindent#1. {\sl Science }{\bf#2}, #3.}
          
\def\lta{\mathrel{\spose{\lower 3pt\hbox{$\mathchar"218$}}
     \raise 2.0pt\hbox{$\mathchar"13C$}}}
\def\gta{\mathrel{\spose{\lower 3pt\hbox{$\mathchar"218$}}
     \raise 2.0pt\hbox{$\mathchar"13E$}}}
     
\def\sec{\hbox{"\hskip-3pt .}}


\font\syvec=cmbsy10                        
\font\gkvec=cmmib10                         

\def\bnabla{\hbox{{\syvec\char114}}}       
\def\balpha{\hbox{{\gkvec\char11}}}	   
\def\bbeta{\hbox{{\gkvec\char12}}}	   
\def\bgamma{\hbox{{\gkvec\char13}}}	   
\def\bdelta{\hbox{{\gkvec\char14}}}	   
\def\bepsilon{\hbox{{\gkvec\char15}}}	   
\def\bzeta{\hbox{{\gkvec\char16}}}	   
\def\boldeta{\hbox{{\gkvec\char17}}}	   
\def\btheta{\hbox{{\gkvec\char18}}}        
\def\biota{\hbox{{\gkvec\char19}}}	   
\def\bkappa{\hbox{{\gkvec\char20}}}	   
\def\blambda{\hbox{{\gkvec\char21}}}	   
\def\bmu{\hbox{{\gkvec\char22}}}	   
\def\bnu{\hbox{{\gkvec\char23}}}	   
\def\bxi{\hbox{{\gkvec\char24}}}           
\def\bpi{\hbox{{\gkvec\char25}}}	   
\def\brho{\hbox{{\gkvec\char26}}}	   
\def\bsigma{\hbox{{\gkvec\char27}}}        
\def\btau{\hbox{{\gkvec\char28}}}	   
\def\bupsilon{\hbox{{\gkvec\char29}}}	   
\def\bphi{\hbox{{\gkvec\char30}}}	   
\def\bchi{\hbox{{\gkvec\char31}}}	   
\def\bpsi{\hbox{{\gkvec\char32}}}	   
\def\bomega{\hbox{{\gkvec\char33}}}        

 
\def\ni{\noindent}
\def\L{{\cal L}}
\def\spose#1{\hbox to 0pt{#1\hss}}
\def\lta{\mathrel{\spose{\lower 3pt\hbox{$\mathchar"218$}}
     \raise 2.0pt\hbox{$\mathchar"13C$}}}
\def\gta{\mathrel{\spose{\lower 3pt\hbox{$\mathchar"218$}}
     \raise 2.0pt\hbox{$\mathchar"13E$}}}
\newcommand{\beq}{\begin{equation}}
\newcommand{\eeq}{\end{equation}}
\newcommand{\lp}{ \left(}
\newcommand{\rp}{ \right)}
\newcommand{\lc}{ \left[}
\newcommand{\rc}{ \right]}
\newcommand{\diff}{{\rm d}}
\newcommand{\cf}{{\it cf.}~}
\newcommand{\ie}{{\it i.e.}~}
\newcommand{\eg}{{\it e.g.}~}
\newcommand{\dtt}{\frac{\partial}{\partial t}}
\newcommand{\drr}{\frac{\partial}{\partial r}}
\newcommand{\dr}[1]{\frac{\partial  #1}{\partial r}}
\newcommand{\beqan}{\begin{eqnarray}}
\newcommand{\eeqan}[1]{\label{#1}\end{eqnarray}}